\documentclass{article}
\usepackage[style=authoryear, giveninits=false,maxcitenames=2, mincitenames=1]{biblatex}
\newcommand{\yearparencite}[1]{(\citeyear{#1})}
\DeclareNameAlias{author}{family}

\title{Timbre Perception, Representation, and its Neuroscientific Exploration: A Comprehensive Review}
\author{Hong Zhang, Jie Lin, Shengxuan Chen}
\date{May 2024}

\begin{document}
\maketitle

\begin{abstract}
Timbre, the sound's unique "color", is fundamental to how we perceive and appreciate music. This review explores the multifaceted world of timbre perception and representation. It begins by tracing the word's origin, offering an intuitive grasp of the concept. Building upon this foundation, the article delves into the complexities of defining and measuring timbre. It then explores the concept and techniques of timbre space, a powerful tool for visualizing how we perceive different timbres. The review further examines recent advancements in timbre manipulation and representation, including the increasingly utilized machine learning techniques. While the underlying neural mechanisms remain partially understood, the article discusses current neuroimaging techniques used to investigate this aspect of perception. Finally, it summarizes key takeaways, identifies promising future research directions, and emphasizes the potential applications of timbre research in music technology, assistive technologies, and our overall understanding of auditory perception.
\end{abstract}
keyword: Timbre perception, Timbre representation, Timbre space, Neural mechanism

\section{Introduction}
For a long time, the conceptual question of what timbre is has been answered with what timbre is not \parencite{bregman_auditory_1994}. Timbre, distinct from pitch and loudness, allows us to distinguish between sound sources \parencite{ansi_1960}. The word, originally French, entered English with three meanings reflecting its evolution in French \parencite{merriam-webster_timbre_nodate}. The first two, referring to a specific drum type and a heraldic crest, are now obsolete. In modern English, timbre solely refers to a sound's quality, independent of pitch, intensity, and loudness.

Interestingly, the French origin hints at its perceptual nature. Cadoz \yearparencite{cadoz_timbre_1987} notes that "timbre" initially referred to a drum with a characteristic "color" sound. Similarly, Scholes \yearparencite{scholes_oxford_1960} defines timbre as "tone quality," encompassing descriptors like "coarse" or "smooth" and color analogies like "scarlet" (trumpet) or "silver" (flute).

When discussing musical timbre, we're essentially talking about perception. Cadoz \yearparencite{cadoz_timbre_1987} reinforces this by stating that timbre can only be understood through human perception. Our auditory system and cognitive abilities play a crucial role in shaping musical expression and conveying emotions through timbre. Timbre's ability to differentiate instruments, create textures, and evoke moods is ultimately determined by how we perceive and interpret sound.

This article delves into the multifaceted world of timbre perception and representation. Section 2 explores how the concept of timbre manifests across various fields, highlighting its significance beyond the realm of music. Section 3 examines the journey from our initial perception of timbre to its representation in different forms. Section 4 ventures beyond acoustics to explore the emerging role of neuroscience in understanding the neural mechanisms underlying our perception of timbre. Finally, the concluding section summarizes key takeaways, identifies exciting future research directions, and emphasizes the potential applications of timbre research in music technology, assistive technologies, and our overall understanding of auditory perception.

\section{Timbre across fields}
\subsection{The Elusive Definition of Timbre}
The question "Whose timbre?" \parencite{smalley_defining_1994} highlights the inherent subjectivity of timbre, making a universally accepted definition challenging \parencite{grey_exploration_1975}. This complexity leads to diverse interpretations across various fields. Musicians, for instance, focus on the expressive potential of timbre, employing terms like "bright" or "warm" to describe a violin's sound or the "breathy" quality of a flute \parencite{kakegawa_effect_2023}. Psychoacoustics, on the other hand, takes a more scientific approach, analyzing the acoustic properties of sound waves to understand how different frequencies and temporal features contribute to our perception of timbre. Cultural backgrounds also influence how we perceive timbre \parencite{kim_timbre_2016}. Certain instruments or sounds may hold specific meanings or evoke particular emotions depending on cultural context \parencite{cowen_what_2020}.

For musicians, pitch, loudness, timbre, and duration are the fundamental pillars of sound. However, timbre is often the most challenging element to describe explicitly. It frequently overlaps with concepts like sound quality and tone color \parencite{slawson_color_1981}, and musicians often employ emotional descriptors.  Musicians use timbre as a powerful tool for expressive nuance in musical composition and performance \parencite{bernays_expressive_2013,de_paula_study_2000, McAdams_perceptual_2019}.  Synesthetic metaphors, like associating a trumpet with "bright red" or a cello with "deep brown" \parencite{scholes_oxford_1960}, further illustrate the effort to capture the perceptual aspects of timbre.

\subsection{Applications and Limitations}
The conventional definition struggles when applied beyond traditional instruments, where the focus is often on replicating the sounds of the real world \parencite{risset_exploration_1999, smalley_defining_1994, krumhansl_why_1989}. In computer music, this limitation is addressed through computer modeling and signal processing techniques. These techniques allow for the creation of entirely new sounds by manipulating and combining acoustic properties in novel ways. This opens up a vast sonic palette for composers and sound designers, pushing the boundaries of musical expression.  For instance, a sound designer might create a unique, otherworldly texture by combining the attack envelope of a plucked string with the high-frequency components of a cymbal.  The ability to analyze and synthesize timbre also empowers musicians to manipulate sounds in real-time during performance, adding a new dimension of control and expression to their music.

\section{From Perception to Representation}
\subsection{A historical review of early approaches}
The 19th century saw the foundation laid for understanding timbre perception. Pioneering researchers like Hermann von Helmholtz \parencite{helmholtz_sensations_1875} proposed that our perception hinges on the prominence of different frequencies within a sound, while Carl Stumpf \yearparencite{stumpf_tonpsychologie_1883} emphasized its subjective nature. The early 20th century brought further advancements. Floyd Watts \yearparencite{watt_psychology_1917} investigated the role of attack transients (initial sound bursts), and Carl Seashore \yearparencite{seashore_manual_1919} developed tests including timbre discrimination tasks.

The latter half of the 20th century witnessed a surge in research on timbre recognition and discrimination. Studies by Preis \yearparencite{preis_attempt_1984}, Wedin and Goude \yearparencite{wedin_dimension_1972}, Miller and Carterette \yearparencite{miller_perceptual_1975}, and Samson et al. \yearparencite{samson_multidimensional_1997} established a strong link between spectral features (the distribution of frequencies in a sound) and perceived musical timbre similarity. These studies primarily used instrument tones as stimuli \parencite{saldanha_timbre_1964,berger_factors_1964,wedin_dimension_1972}, however, Luce's \yearparencite{luce_physical_1963} work employed computer and analog-to-digital equipment for spectral analysis, a pioneering use of technology in this field.

The concept of categorical perception, where listeners perceive sounds as belonging to distinct categories rather than a continuum, became a focus. Studies by Cutting and Rosner \yearparencite{cutting_categories_1974-1} and Grey \yearparencite{grey_exploration_1975} explored this in musical sounds. Additionally, research by Bismarck \yearparencite{von_bismarck_timbre_1974} investigated verbal descriptions and evaluations of timbre, while Grey \yearparencite{grey_exploration_1975} explored the discrimination of musical tones.

A significant limitation of these studies was their reliance on isolated tones, devoid of any musical context. Grey's \yearparencite{grey_timbre_1978} groundbreaking research addressed this by testing three instrumental timbres (clarinet, trumpet, and bassoon) in various musical contexts. His findings revealed that the context significantly impacted the types of timbre differences listeners perceived. Musical contexts amplified spectral differences, while isolated contexts allowed for clearer comparison of temporal details (how sound unfolds over time). This suggests, as Grey noted, that different factors influence timbre discrimination, and their importance varies depending on the context.

Grey's \yearparencite{grey_exploration_1975} work employed computer analysis and synthesis to delve deeper into the complexity of musical timbre. He manipulated sound properties to synthesize natural instrument tones from simplified components. These simplified tones, valuable for further psychoacoustic studies, served to reduce the number of factors influencing perception. Notably, Grey identified the attack segment as a crucial cue for instrument recognition, highlighting the multifaceted nature of timbre perception, which involves both steady-state characteristics and transient features. This work paved the way for future research in timbre representation.

\subsection{Exploring Timbre representation with MDS}

The intricate nature of timbre demanded novel approaches to its exploration and analysis. Progress relied heavily on advancements in digital technology, data analysis techniques, and computer algorithms. Pioneering this application in 1973, Wessel revealed the crucial role of both spectral and temporal features in timbre perception. He examined the concept of timbre space which initially introduced by Plomp \yearparencite{plomp_timbre_1970}, and explored its use with Multidimensional Scaling (MDS) \parencite{shepard_analysis_1962} techniques. Timbre space serves as a geometric model, with similar tones positioned closer together, representing their perceptual dissimilarities. Wessel explored its potential as a navigational tool for composers and conducted an MDS experiment on perceived dissimilarities of orchestral instruments. This research shed light on the relationship between acoustic properties and timbre perception.

Multidimensional Scaling is a powerful statistical technique that has been instrumental in mapping out the hidden dimensions of timbre space. Prior to MDS, timbre research was limited by its reliance on predefined acoustic features, potentially overlooking aspects of human perception. MDS offered a paradigm shift. Developed by Shepard in 1962, it leverages direct dissimilarity ratings from listeners. Instead of imposing pre-conceived notions about relevant features, listeners judge how different pairs of sounds seem to them. This approach allows researchers to capture the essence of how humans perceive timbre differences.

The core of MDS lies in its ability to translate these dissimilarity judgments into a spatial representation. By analyzing the ratings, MDS creates a timbre space, a multidimensional map where each point represents a specific sound. Sounds perceived as similar by listeners are positioned closer together in this space, while highly dissimilar sounds occupy more distant locations. The number of dimensions in the space is determined by the complexity of the data and the underlying perceptual features.

In essence, MDS offers significant advantages for timbre research. First, it's data-driven, relying on listener judgments rather than pre-determined acoustic features. This captures the full spectrum of perceived timbral differences. Second, MDS provides a visual representation (timbre space) of sound relationships. While interpreting the extracted dimensions requires additional psychoacoustic knowledge, and the quality of results relies on listener ratings, MDS remains a foundation of timbre research.

\subsection{Milestones in the evolution of timbre space}

Early explorations of timbre space, as seen in Wessel's \yearparencite{wessel_psychoacoustics_1973} work employing MDS identified two key perceptual dimensions. Listeners' judgments are analyzed by a program called INDSCAL that creates a two-dimensional space where the location of a sound reflects its perceived similarity to others (see figure 1 in \cite{wessel_psychoacoustics_1973}). The vertical dimension likely relates to the steady-state portion of a sound's spectrum, while the horizontal dimension might capture the influence of the initial attack. This "timbre space" allows researchers and composers to see how different sounds relate to each other perceptually, and the paper acknowledges the potential for a three-dimensional space to capture even finer details of timbre perception.

Building on these findings, Grey \yearparencite{grey_exploration_1975} employed synthetic orchestral tones and listener judgments analyzed via MDS. This approach yielded a three-dimensional timbre space (see figure 11 in \cite{grey_exploration_1975}). Similar to Wessel's results, the first dimension again reflected spectral energy distribution. The second dimension captured attack rapidity, but also interestingly correlated with spectral fluctuation (changes in spectral content over time). The third dimension represented the spectral balance during the attack transient, highlighting how the initial energy distribution varied across frequencies. These findings underscore the importance of both spectral and temporal characteristics in our perception of timbre.

Traditional MDS techniques aim for lower-dimensional representations while preserving distances between data points. However, given the complexities of timbre, the adequacy of a two- or three-dimensional space as a comprehensive representation model is brought into question. Researchers like Krumhansl \yearparencite{krumhansl_why_1989} and McAdams et al. \yearparencite{mcadams_perceptual_1995} investigated the concept of "timbre specificity," unique perceptual characteristics that contribute to identifying individual timbres. These specificities can arise from the mechanical design of the instrument, such as the harpsichord's action or the clarinet's air column shape \parencite{Donnadieu_mental_2007}.

McAdams et al. \yearparencite{mcadams_perceptual_1995} employed an extended version of MDS in a study using synthesized musical tones. Their analysis aimed to estimate several factors, including the number of listener groups, the location of each timbre on common dimensions, and the specificities of each timbre. They also attempted to qualitatively interpret the specificities, identifying both continuous features (e.g. raspiness of attack) and discrete features (e.g. high-frequency transients). These specificities could account for additional continuous dimensions and discrete features of varying perceptual salience. The study’s psychophysical interpretation of the common dimensions, informed by the acoustic parameters outlined by Krimphoff et al. \yearparencite{krimphoff_caracterisation_1994}, identified them as quantifiable through log-rise time, spectral centroid, and spectral flux. The results suggested that musical timbres possess specific attributes not captured by these shared perceptual dimensions. Furthermore, the study provided valuable insights into multidimensional analysis, including selecting the appropriate spatial model, determining the number of dimensions, including specificities, and estimating class belongingness based on factors like musical background.

While adding more dimensions might seem intuitive for representing the intricacies of musical sound, Pollard and Jansson's \yearparencite{pollard_tristimulus_1982} study presented a distinct approach. Recognizing the limitations of one-dimensional scales for multidimensional timbre, they proposed a “tristimulus” method focusing on the graphical representation of a sound's time-dependent timbre behavior. This method analyzes the sound through filters, capturing the loudness of specific frequency bands and their evolution over time. The resulting data is plotted on a tristimulus diagram (see figure 3, 6, 7 in \cite{pollard_tristimulus_1982}), offering a simplified yet informative visual representation of the sound's timbre throughout its duration. 

Pollard and Jansson's tristimulus method, inspired by color vision, offers a novel way to visualize dynamic timbre changes. While intuitive for understanding attack-to-steady-state transitions, it focuses on limited spectral features, potentially neglecting other key aspects of timbre like inharmonicity and temporal envelopes. This selectivity may limit its effectiveness, particularly for complex sounds or subtle variations. Additionally, interpreting the tristimulus diagram requires expertise, introducing potential subjectivity and variability in analysis. Despite these limitations, tristimulus method offers a valuable contribution to the field by highlighting the importance of considering dynamic changes in timbre. It provides a practical tool for visualizing the sound transient, which is a critical aspect of musical timbre often overlooked in earlier techniques.

\subsection{Higher dimensional representations}
High-dimensional audio representations have become a staple in the field of speech processing, with ongoing discussions centered around the most effective degree of detail required for accurate modeling. In Dau et al’s \yearparencite{dau_modeling_1997} study, they employed a temporal modulation filter bank model, which, while effective, did not explicitly incorporate spectrotemporal modulations. Elhilali et al. \yearparencite{ELHILALI2003331} incorporated a wider range of spectrotemporal modulations, significantly improving speech intelligibility prediction.

Within the realm of music information retrieval, the trend observed is the application of a vast array of meticulously crafted audio descriptors, as exemplified by the work of Siedenburg et al. \yearparencite{siedenburg_comparison_2016}. The abundance of descriptors in music information retrieval led Peeters et al. \yearparencite{peeters_timbre_2011} to identify redundancies, prompting a reevaluation of their necessity. In alignment with this, Patil et al. \yearparencite{patil_music_2012} and Thoret et al. \yearparencite{thoret_perceptually_2017} showcased the efficacy of spectrotemporal modulation features in achieving robust automatic classification of musical instruments, underscoring the viability of these features in timbre analysis. McDermott and Simoncelli \yearparencite{mcdermott_sound_2011} delved into the nuances of sound texture perception, employing an analysis-resynthesis methodology to assess the perceptual fidelity of re-created sounds. Their findings indicated that relying solely on the statistics of individual frequency channels was inadequate for generating lifelike textures, thereby spotlighting the critical role of inter-channel correlations.

\subsection{Novel approaches for timbre manipulation}
The 21st century has witnessed significant strides in the understanding and manipulation of musical timbre, leading to transformative approaches in sound synthesis and musical expression. The work of Bitton et al. \yearparencite{bitton_vector-quantized_2020} stands out in this context, presenting an auto-encoder architecture that disentangles loudness from other sound features, thereby enabling more precise control over timbre during sound synthesis. This novel approach has opened new avenues for transforming sounds between instruments and vocals, offering musicians a greater palette of sonic possibilities. The method's versatility is particularly impactful for music production, where it can facilitate the creation of unique soundscapes that were previously unattainable. 

Complementing this practical development, Vahidi et al. \yearparencite{vahidi_timbre_2020} investigated how specific acoustic properties of synthesized sounds from subtractive synthesizers influence their perceived timbre, guiding the design of more intuitive synthesizers. By analyzing listener dissimilarity ratings, this study sheds light on how specific sound features contribute to our perception of timbre, guiding the design of future synthesizers with more intuitive controls and a richer sonic variety.

In a similar vein, Sköld \yearparencite{skold_visual_2022} has tackled the challenge of visually representing timbre in musical notation. He innovates by integrating perception-based symbols into a staff notation system to depict spectral qualities visually (width, centroid, density). This system, tested on electroacoustic works (see figure 7-9 in \cite{skold_visual_2022}), is particularly effective for pieces with multiple sounds. Building on the potential for practical application, Sköld acknowledges the inherent complexity of timbre. However, Sköld argues that his system goes beyond a simple notation tool. It empowers musicians by integrating timbre notation with established musical parameters, fostering imagination, aiding perception of the sonic landscape, and facilitating communication through a practical framework for incorporating timbre into scores.

The creative potential of timbre is further explored by Caillon et al. \yearparencite{caillon_timbre_2020}, who have utilized machine learning models to learn and control loudness-independent sound features. Their work aligns with the concept of multidimensional timbre spaces, underscoring the importance of training models for specific functionalities. This research not only paves the way for the development of user interfaces that allow for exploration of these latent dimensions but also enables musicians and sound designers to engage in novel sonic exploration and expression.

\subsection{Advancing Through Machine Learning Techniques}
The exploration of timbre perception has seen exciting advancements in recent years. Machine learning algorithms, with their ability to automatically learn patterns from data, have been widely adopted for analyzing audio signals \parencite{paterna_timbre_2017,hernandez-olivan_comparison_2021} with applications in developing real-time manipulation tools for musicians \parencite{ganis_real-time_2021} and creating comprehensive models of human auditory perception \parencite{jiang_analysis_2020}. Techniques like adversarial training and generative models further show promise for enhancing timbre representation quality \parencite{kim_adversarial_2022}.

While machine learning offers intriguing possibilities for future research, a solid understanding of traditional acoustic principles remains essential. This understanding provides a foundation upon which more complex models and techniques can be built. By integrating these established approaches with the power of machine learning, we can advance our understanding of timbre perception to even greater heights. However, the exploration of timbre perception extends beyond the realm of acoustics. The human brain plays a crucial role in how we perceive and interpret sound. In the next section, we will delve into the exciting field of neuroscience, examining how our brains process and decode the complex information embedded within a sound's acoustic properties.

\section{Beyond Acoustics: Neuroscience-based Exploration of Timbre Perception}
\subsection{Brain Regions Specialized for Timbre Processing}
The utilization of advanced magnetic resonance imaging techniques, including functional Magnetic Resonance Imaging (fMRI) and Positron Emission Computed Tomography (PET), has been instrumental in uncovering the neural substrates of timbre perception. 

Several studies \parencite{Zatorre_Cortex_2001,Hall_Cortex_2002} used PET and fMRI to monitor brain activity while participants were presented with various tones, including pure tones, harmonic tones, and those with varying modulations. These studies reveal that the core auditory cortex is sensitive to temporal variations, with a left hemisphere bias. In contrast, the anterior superior temporal regions respond bilaterally to spectral variations, with a rightward inclination. This pattern suggests a complementary specialization of the hemispheres, with the right hemisphere processing spectral information and the left handling temporal information. These findings align with previous work on the lateralization of brain function in timbre perception \parencite{robin_auditory_1990,platel_structural_1997}.

Furthermore, the research \parencite{Zatorre_Cortex_2001} supports the hierarchical model of auditory processing, where core areas handle initial perception, and belt cortical areas are involved in higher-order processing of complex sounds. Interestingly, the studies suggest that complex sounds, characterized by a rich spectrum of frequencies and changing pitches, elicit a stronger response in the brain's auditory regions.

Specifically, when participants listened to harmonic tones (multi-frequency sounds with a richer musical quality), the right Heschl's gyrus and the lateral supratemporal plane on both sides showed increased activity compared to single-frequency tones \parencite{Hall_Cortex_2002}. This highlights the brain's heightened sensitivity to sounds with a broader frequency range. Additionally, frequency-modulated tones, which change pitch over time, provoked more pronounced activation in these areas, along with the left Heschl's gyrus and superior temporal sulcus, compared to constant-pitch sounds. This indicates a particular sensitivity within the brain's auditory regions to changes in both the spectral and temporal dimensions of sound.

The research conducted by Platel et al. \yearparencite{platel_structural_1997} supports the notion of functional independence among different musical elements, with distinct neural networks activated for pitch, rhythm, and timbre.The left hemisphere's involvement in tasks associated with musical familiarity and rhythm processing underscores its role in semantic processing and music recognition. 

Moreover, brainstem's distinctive responses to frequently-practiced musical instruments, as observed in Strait et al.'s \yearparencite{strait_specialization_2012} study, further exemplifies the complexity of timbre perception. The results indicated that musician’s auditory brainstems responded most accurately to the timbre of their own instrument, suggesting that extensive musical practice refines subcortical sound processing, leading to a more acute perception of musical timbre. Collectively, these findings highlight the intricate interplay of neural networks that facilitate our perception and appreciation of the rich tapestry of musical sounds, emphasizing the significance of hemispheric specialization and cognitive strategies in our auditory experience of music.

\subsection{A new aspect for timbre brightness perception}
The complexity of timbre perception extends beyond the identification of specific instruments. Timbre also encompasses characteristics like brightness, which influences our emotional response to music. Maxwell et al.’s \yearparencite{maxwell_new_2021} study introduces a novel auditory theory positing that neural fluctuations (low-frequency variations in the auditory nerve's activity) are pivotal in the perception of timbre brightness. These fluctuations are not merely a reflection of external sounds but can also arise and be modified within the auditory system itself. The study proposes that a process known as peak sharpening occurs in the early auditory system, between the auditory nerve and the midbrain. This process refines the neural representation of spectral peaks, rendering them more precise and distinct, which is crucial for the perception of musical timbre. The research suggests that the center of mass of the most salient spectral peaks, as encoded in the midbrain, constitutes the neural representation of brightness. This implies that the concept of brightness in musical timbre may be intrinsically linked to the presence of locally prominent spectral peaks and temporal modulation. Intriguingly, the study discovered that the addition of low-level pink noise could enhance the neural representation of spectral peaks by amplifying neural fluctuations at tonotopic locations adjacent to and beyond the spectral peaks. This enhancement sharpens the representation and could potentially refine the perception of timbre.

\subsection{Embodied Perception of Timbre}
Timbre processing has been associated with the sensorimotor area, indicating a multifaceted neural engagement that transcends basic auditory stimulus processing. Wallmark et al.’s \yearparencite{wallmark_embodied_2018} study, conducted through three distinct experiments, has investigated the embodied aspects of timbre perception. The findings suggest a motor component in timbre processing, particularly for 'noisy' timbral qualities, hinting at a possible enactive mechanism. As participants' aversion to a timbre intensifies, increased activity in somatomotor areas, insula, and the limbic system is observed, alongside a diminished connectivity between the premotor cortex and insula. These results bolster theories of embodied music cognition and the role of timbre in shaping emotional responses to music.

Another study by Alluri et al. \yearparencite{alluri_large-scale_2012} revealed that music listening activates an extensive array of regions, encompassing those related to cognition, motor function, and emotions. Timbre-related acoustic elements were linked to activations in the bilateral superior temporal gyrus, which plays a role in sensory and default mode network cerebrocortical areas, as well as cognitive regions of the cerebellum. Notably, the superior temporal gyrus, cerebellum, and specific areas of the right hemisphere were particularly active in processing timbral features.

\subsection{The Emotional Power of Timbre}
Timbre's influence extends beyond shaping our auditory experience. Hailstone et al.’s study \yearparencite{hailstone_its_2009} have shown that timbre affects the perception of emotion in music, its precise contribution remains elusive due to the challenge of isolating it from other musical parameters. This elusiveness may be central to its impact. Research \parencite{spreckelmeyer_preattentive_2013} employing EEG technology suggests that the brain can automatically and preattentively process emotional information in music, particularly through the timbre of the sounds. 

Spreckelmeyer et al.’s \yearparencite{spreckelmeyer_preattentive_2013} ERP (event-related potential) experiment identified a mismatch negativity (MMN) for emotional deviants, indicating that the brain can form an emotional memory trace at a preattentive level.  ERP is a technique that measures the electrical activity of the brain in response to specific stimuli, and MMN is a specific component of the ERP waveform that reflects the brain's detection of a change from a repeated sound. This suggests that the brain's capacity to process emotional information in music is automatic and preattentive, primarily through the timbre of the sounds. Expanding on prior findings, this research demonstrates that emotional deviance in music can be detected without attention and that the neural mechanisms involved are analogous to those observed in the processing of other emotional stimuli. These insights contribute to our understanding of how the brain rapidly decodes emotional content in music, with implications for both the neuroscience of emotion and the psychology of music perception. In essence, despite its subtlety, timbre serves as a potent, yet often overlooked, element in forging our emotional bond with music.

In essence, neuroscience paints a fascinating picture of timbre perception. Distinct brain regions specialize in processing the spectral and temporal cues that define timbre. Musical practice can refine this perception, and timbre characteristics like brightness influence our emotional responses. Notably, the brain appears to automatically process emotional information conveyed through timbre. These findings highlight the intricate interplay between brain networks, cognition, and our emotional experience of music, underscoring the multifaceted nature of timbre perception.

\section{Conclusion}
The study of musical timbre perception and representation has made significant strides in recent years, offering valuable insights into the complex nature of how we perceive and process the unique 'color' of sounds. The definition and understanding of timbre have evolved from early psychoacoustic models to the sophisticated multidimensional scaling (MDS) techniques that have mapped the perceptual space of timbre with greater precision. These advancements have not only enhanced our appreciation of the role of timbre in music but have also paved the way for more nuanced digital representations of sound.

Despite these gains, the underlying neural mechanisms of musical timbre perception remain only partially understood. The current research has its merits, with neuroimaging studies such as fMRI and PET scans providing a window into the brain's response to timbre. However, these methods also have limitations, including the challenge of capturing the dynamic and real-time nature of auditory perception. Moreover, the focus on the auditory cortex and other brain regions involved in sound processing does not fully account for the complex interplay between sensory input, cognitive processing, and emotional response that characterizes our experience of music.

Several questions remain open for future research. For instance, how do individual differences in musical training and cultural background influence timbre perception? What is the role of attention and expectation in the auditory scene analysis that leads to timbre differentiation? How can we integrate the knowledge of timbre perception into the development of new musical interfaces and assistive technologies for individuals with auditory disabilities?

In terms of future directions, there is a need for more interdisciplinary research that combines cognitive psychology, neuroscience, and musicology. This could involve the development of new experimental paradigms that capture the online aspects of timbre processing and the use of machine learning algorithms to model the complex patterns of neural activity associated with timbre perception. Additionally, there is a call for research that explores the therapeutic potential of timbre in music therapy and its application in the rehabilitation of auditory processing disorders.

In conclusion, while we have made progress in understanding the perception and representation of musical timbre, there is still much to discover. The journey ahead is exciting, with the potential to unlock new dimensions of auditory experience and to develop innovative applications that can enhance both our scientific understanding and our musical practices. The continued pursuit of knowledge in this area will undoubtedly enrich our comprehension of the human auditory system and its remarkable capacity to perceive and appreciate the diverse tapestry of sounds that make up our musical world.

\end{document}